\documentclass[runningheads]{llncs}
\usepackage[T1]{fontenc}
\usepackage{graphicx}
\usepackage[numbers]{natbib}
\usepackage{amsmath}
\usepackage{algorithm}
\usepackage{algpseudocode} 
\usepackage{flushend}

\usepackage{graphicx}
\usepackage{subcaption}
\usepackage{wrapfig}
\usepackage{cleveref}
\begin{document}
\title{Enhancing Web Spam Detection through a Blockchain-Enabled Crowdsourcing Mechanism}
\titlerunning{Crowdsourced Spam Detection}
%
\author{Noah Kader\orcidID{0009-0005-9172-1106} \and
Inwon Kang\orcidID{0000-0001-8912-287X} \and
\\Oshani Seneviratne\orcidID{0000-0001-8518-917X}\thanks{corresponding author}
}
\authorrunning{Kader, Kang, and Seneviratne}
\institute{Rensselaer Polytechnic Institute, Troy, New York, USA\\
\email{noahkader@gmail.com,kangi@rpi.edu, and senevo@rpi.edu}}
\maketitle              
\begin{abstract}


The proliferation of spam on the Web has necessitated the development of machine learning models to automate their detection. However, the dynamic nature of spam and the sophisticated evasion techniques employed by spammers often lead to low accuracy in these models. Traditional machine-learning approaches struggle to keep pace with spammers' constantly evolving tactics, resulting in a persistent challenge to maintain high detection rates. To address this, we propose blockchain-enabled incentivized crowdsourcing as a novel solution to enhance spam detection systems. We create an incentive mechanism for data collection and labeling by leveraging blockchain's decentralized and transparent framework. Contributors are rewarded for accurate labels and penalized for inaccuracies, ensuring high-quality data. A smart contract governs the submission and evaluation process, with participants staking cryptocurrency as collateral to guarantee integrity. Simulations show that incentivized crowdsourcing improves data quality, leading to more effective machine-learning models for spam detection. This approach offers a scalable and adaptable solution to the challenges of traditional methods.

\end{abstract}
%
%
%
%
%
%

\section{Introduction}

Web spam broadly refers to deceptive or malicious techniques to manipulate search engine rankings, deceive users, or engage in illicit online activities. Web spammers aim to deceive search engines and users into perceiving their content as valuable or legitimate, even though it may be of low quality or carry malicious intent~\cite{spirin2012survey}. 
Web spammers frequently use targeted keywords excessively in content or metadata to manipulate search engine rankings.
This practice artificially inflates a web page's relevance for specific search queries, even if the content is not genuinely valuable~\cite{gyongyi2005web}. 
For instance, "doorway," "bridge," or "gateway" pages are low-quality web pages optimized to rank well for specific search queries~\cite{chellapilla2007taxonomy}. They often provide users with little or no valuable content but act as entry points to redirect visitors to other websites, including spam or malicious sites. 
Spammers often copy or "scrape" content from legitimate websites and republish it on their sites. This technique aims to deceive search engines into considering the copied content as original, leading to undeserved high rankings and potentially stealing traffic from the original source~\cite{ntoulas2006detecting}. 

In the context of online advertising, web spammers engage in click fraud to generate illegitimate clicks on pay-per-click ads, which can be done using automated bots or by incentivizing individuals to click on ads, aiming to drain advertisers' budgets or artificially increase ad impressions and click-through rates~\cite{dave2012measuring}.
Spammers may also hide keywords or links by using techniques such as setting the text color to match the background, positioning them off-screen, or using tiny font sizes. These hidden elements are intended to deceive search engines by including additional keywords~\cite{urvoy2008tracking}. There are also various tactics to manipulate the number and quality of inbound links to spam websites. These tactics include buying or exchanging links, participating in link farms or exchange networks, and using automated programs to generate large volumes of low-quality or irrelevant backlinks~\cite{wu2005identifying,gyongyi2005link}. 

Web spam techniques continue to evolve as search engines and security systems enhance their detection methods. Search engine algorithms and spam detection systems employ sophisticated techniques to identify and penalize web spam, striving to provide users with high-quality and trustworthy search results~\cite{spirin2012survey,crawford2015survey}.
However, these models often encounter difficulties due to the dynamic nature of spam websites and spammers' utilization of sophisticated evasion techniques. Consequently, there is a critical demand for innovative approaches to enhance web spam detection effectiveness. 

Crowdsourcing could be a viable alternative to this problem.  
It involves obtaining input or data by enlisting the services of a large number of people, typically from an online community~\cite{howe2009crowdsourcing}. It has been successfully applied in various domains, including image recognition, language translation, and data labeling.
In the context of web spam detection, crowdsourcing can leverage the collective intelligence and diverse perspectives of numerous contributors to identify and label spam content accurately. 

Blockchain, first introduced by \citet{nakamoto2008bitcoin}, has been widely applied in finance, supply chain management, and data security. It has also been explored as an incentive mechanism in various fields. However, our work makes a unique contribution by integrating blockchain into the realm of web spam detection, where accurate, high-quality data is crucial for improving model performance. 
This approach adds significant value by expanding blockchain's applicability to AI-driven web spam detection, an area that relies heavily on trustworthy data to enhance model accuracy and performance.

\subsection{Contributions}

We propose a mechanism that incentivizes users to contribute high-quality data, thereby improving the overall accuracy of our web spam detection models. 
Building on prior research~\citet{xu2022improving,Harris_2019}, we introduce a blockchain-based incentive mechanism that aligns with users' needs and ensures data integrity and quality.
Our objective is to create an effective incentive mechanism that encourages users to provide accurate and valuable data and discourages the submission of inaccurate or malicious information by untrustworthy actors.


Our contributions include (1) a blockchain-based incentive mechanism, (2) the demonstration of its efficacy in web spam detection, and (3) the potential applicability to other domains requiring high-quality crowdsourced data.

\section{Related Work}

Many studies have explored blockchain integration in machine learning~\cite{seneviratne2022blockchain}, but our work uniquely combines this with incentivized crowdsourcing to address web spam detection.
In this section, we outline the state of the art in AI for spam detection and blockchain-enabled crowdsourcing mechanisms and compare and contrast them with our work.

\subsection{Spam Detection}

Much of the recent work in spam detection is related to email rather than web spam detection.
\citet{farooq2019survey} presents a comprehensive survey of techniques commonly used by spam websites, serving as a valuable reference for our work. 
\citet{sahmoud2022spam} utilized BERT for spam detection, demonstrating the effectiveness of transformer models in understanding context and semantics. 
\citet{kim2020deepcapture} proposed "DeepCapture," a deep learning and data augmentation approach for image spam detection. 
These works provide valuable insights and complement our approach by addressing different aspects of spam detection with advanced methodologies, which we leveraged in our spam detection algorithm.

Spam detection is especially challenging in a setting with limited resources, such as a blockchain-based environment. 
In this regard, past works have found efficient features that can be extracted from the source HTML of a website's domain URL and have shown good performance.
\citet{jelodar2017systematic} propose a framework to learn and improve features using regex pattern matching. The proposed approach is iterative, where the features are first encoded and used to train the model, which then finds better features while improving accuracy. 
While this approach demonstrates the effectiveness of keyword matching, its complexity may hinder its practicality. 
\citet{ntoulas2006detecting} explored various web page features for spam detection, which we extend by integrating a blockchain-based incentive system to enhance feature set evolution and accuracy.
These include repeated mentions of keywords and visible elements by styling them appropriately using Cascading Styling Sheets (CSS) or the number of anchor texts (text links to other pages). The authors showcase a high performance of over 90\% precision accuracy in detecting spam with proposed features and a decision tree classifier.
\citet{markines2009social} define a specific spam problem called "social spam" and define several associated features, such as styling features or the number of links on a page, and show their performance of around 97\% 

\citet{mamun2016detecting} contribute a new dataset (ISCX-URL2017)\footnote{https://www.unb.ca/cic/datasets/url-2016.html} on spam data on the web. The authors also make use of network features to analyze this dataset.
While the network features are irrelevant to our work, since this dataset is much newer, we created our feature set using the URLs available in this dataset.
We also build a low-computational spam filtering application using features from these past works.

\subsection{Blockchain-Enabled Crowdsourcing}

\citet{harvey2018blockchain} discuss the advantages of using blockchains, where the authors argue that blockchain technology offers near-zero transaction cost, transparency, and permissioned identity sharing. These benefits can be applied to personalized spam detection, as users can share their unique identity with smart contracts and engage in data exchange at a minimal cost. The technology has several benefits, such as automation and connection to sophisticated machine-learning techniques through oracles~\cite{pisano2023predictchain}.
However, improvements can be made to enhance scalability and efficiency, ensuring the system can handle many user interactions without compromising performance.

\citet{sheikh2021cryptocurrency} propose a blockchain-based email system to discourage spammers. By adding a small fee to every email, spammers are disincentivized from sending large batches of emails due to the high total cost.
While this approach provides a disincentive for mass spamming, scalability remains a concern, as each email triggers a transaction on the blockchain. Optimizations should be explored to minimize transactional overhead and ensure the system can handle a high volume of email transactions effectively to improve this method.
Similarly, \citet{choudhari2021spam} propose a blockchain-based email system where transactions are attached to each email, and the recipient decides whether the email is spam or not, determining the refund of the transaction. If a wallet has been tagged as spam, the system will show the following emails from the address associated with the wallet as spam. While this approach can help identify spammers, it relies on individual recipient judgments, which may introduce biases or subjective evaluations. Enhancements can be made to incorporate more sophisticated spam detection techniques or community-based voting systems to ensure more reliable spam identification, which is the focus of our work in this paper.

\citet{nguyen2019privacy} propose a blockchain-based geo-marketplace where users can sell arbitrary data tagged with their geo-location. The authors propose a system in which the users can list their data in a marketplace curated by a trusted authority, which allows buyers to search for and buy the data. Spammers are filtered by posting the hashed location values on the blockchain before their data is listed. Malicious actors cannot modify the location data associated with the sale because the data cannot be modified once posted on the blockchain. 
The use of blockchain ensures the integrity and immutability of location data. However, the scalability and efficiency of the marketplace can be improved by exploring mechanisms to handle a large number of data listings and transactions while maintaining low costs and fast processing times.

\citet{xu2022blockchain} propose a blockchain-based crowdsourcing system that could help alleviate the data shortage problem in machine learning by incentivizing data providers to contribute data to the system. 
\citet{kadadha2022chain} propose a machine-learning model for behavior prediction in blockchain-based crowdsourcing systems. They argue that utilizing blockchain transaction data could create a more accurate and trustworthy model for predicting behavior in such systems. The authors presented a detailed architecture of their proposed model, which utilized a combination of supervised and unsupervised learning techniques and evaluated its performance through simulations.
\citet{Harris_2019} also showcase the decentralization and collaboration aspects of blockchain technology and their potential to enhance machine learning. They proposed a blockchain-based collaborative machine learning framework that enables multiple parties to contribute their data and models, collaborate on model training, and share the results.
We drew inspiration from both these works in developing our targeted incentivized web spam detection system.

\section{Methodology}
\label{sec:methodlogy}

\subsection{System Overview}


Our system leverages blockchain technology to incentivize high-quality data submissions for web spam detection. The primary components include a smart contract facilitating a crowdsourcing platform and a machine-learning model running on an off-chain oracle.

\begin{wrapfigure}{r}{0.6\textwidth}
  \centering
  \vspace{-5ex}
  \includegraphics[width=\linewidth]{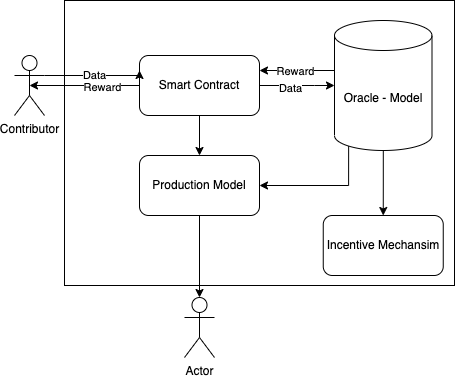}
  \caption{System Overview}
  \vspace{-3ex}
  \label{fig:mech_overview}
\end{wrapfigure}

The system overview, depicted in \Cref{fig:mech_overview}, illustrates the flow of operations. Contributors interact with a smart contract deployed on the blockchain, submitting their data and staking their currency. The smart contract then forwards the data to the web spam model outlined in \Cref{sec:prediction}, which processes the data. 
The contributors are compensated based on how much the contributed data agrees with the base model accuracy.

The proposed mechanism involves contributors who submit data to the model, with the potential to earn profits based on the quality of their submissions. To ensure the system's integrity, we classify contributors into \textit{Good Actors} and \textit{Bad Actors}. 
\textit{Good Actors} are contributors who operate in good faith. While they are not restricted to submitting only non-spam websites, they must ensure they have accurate labels. On the other hand, \textit{Bad Actors} aim to disrupt the model's integrity or exploit the mechanism for undeserved rewards by intentionally providing incorrect labels or repeatedly submitting the same data point that has already triggered rewards.

\subsection{Prediction Model}
\label{sec:prediction}

Our implementation utilizes a machine learning model trained on a labeled spam and non-spam instances dataset. 
Our spam detection model's architecture is modular and can be adapted to incorporate state-of-the-art (SotA) algorithms as they become available.

\subsubsection{Training and Evaluation}

The model is trained on a labeled dataset containing examples of both spam and non-spam instances. During training, it learns to optimize its parameters to minimize classification errors. Evaluation metrics include accuracy in assessing the model's performance. 

\subsubsection{Weight Calculation Algorithm}

The weight assigned to each contributor is crucial for determining the rewards. We used the following algorithm to calculate the weight:

\begin{algorithm}
\caption{Weight Calculation with Penalization for Repeated Submissions}
\begin{algorithmic}[1] 
\Procedure{CalculateWeight}{newData, baseModel, submissionHistory}
    \State accuracy\_base $\gets$ Evaluate(baseModel, baseDataset)
    \State $combinedData \gets baseDataset \cup newData$
    \State newModel $\gets$ Train(baseModel, combinedData)
    \State accuracy\_new $\gets$ Evaluate(newModel, baseDataset)
    \State $baseWeight \gets \frac{(accuracy\_new - accuracy\_base)}{accuracy\_base}$
    
    \State $penaltyFactor \gets 1.0$
    \For{dataPoint \textbf{in} newData}
        \If{dataPoint \textbf{in} submissionHistory}
            \State $penaltyFactor \gets penaltyFactor \times 0.9$ \Comment{Reduce weight by 10\% for each repeated data point}
        \EndIf
    \EndFor

    \State $finalWeight \gets baseWeight \times penaltyFactor$
    
    \State \Return finalWeight
\EndProcedure
\end{algorithmic}
\label{alg:weight}
\end{algorithm}

The weight calculation algorithm determines the contribution weight of new data submissions while penalizing repeated submissions to ensure data quality and diversity. Initially, the base model's accuracy is evaluated using the existing dataset. The new data is then combined with the base dataset, and the model is retrained with this combined data. The new model's accuracy is evaluated again to compute the base weight, which is defined as the relative improvement in accuracy compared to the base model.

To account for repeated submissions, the algorithm maintains a submission history, tracking previously submitted data points. A penalty factor is initialized to 1.0 and is progressively reduced by 10\% for each data point in the new submission that has been previously submitted. This reduction ensures that the final weight reflects the novelty and uniqueness of the data contributed. The final weight is computed by multiplying the base weight by the penalty factor, and this penalized weight is returned as the algorithm's output. This approach encourages contributors to submit unique and valuable data, enhancing the overall quality of the dataset.

If the weight is positive, the new data will be added to the production model dataset, and the contributor will receive a reward. 
Conversely, if the weight is negative, the new data will be considered bad, and the contributor will lose their stake. 

\subsection{Model Deployment}

\subsubsection{Oracle}
The model is not stored on the blockchain; we utilize an off-chain oracle~\cite{muhlberger2020foundational}. Running machine-learning models on-chain is cost-prohibitive due to the required computational resources, translating into high gas fees on blockchain platforms such as Ethereum, our blockchain platform of choice. Moreover, smart contract size limitations and the computational constraints inherent in Ethereum further restrict the feasibility of deploying complex machine-learning models directly on the blockchain. These limitations hinder the on-chain execution of resource-intensive tasks such as model training and inference.
By leveraging an off-chain oracle, we can perform the heavy computational lifting off-chain and subsequently relay the results back to the blockchain securely and verifiably. This approach mitigates the cost concerns and circumvents the storage and processing constraints of blockchain environments. The off-chain oracle serves as a bridge between the blockchain and external data sources or computational services, enabling efficient and scalable execution of machine-learning models without compromising the integrity and transparency of the blockchain.
Furthermore, the use of off-chain oracles enhances our system's flexibility and scalability. Off-chain computation allows for the integration of advanced machine learning techniques and algorithms that would otherwise be impractical to implement on-chain. This separation of concerns ensures that the blockchain maintains its primary role in ensuring security, immutability, and decentralized consensus. At the same time, the computationally intensive tasks are handled off-chain for efficiency and cost-effective purposes.

\subsubsection{Deployment Strategy}

Our deployment strategy involves using smart contracts to manage data submissions, stakes, and rewards. 
Smart contracts are self-executing contracts with the terms of the agreement directly written into code~\cite{buterin2014next}. They automatically enforce and execute the terms of a contract when predefined conditions are met. Smart contracts run on blockchain technology, ensuring transparency, immutability, and security. These properties make smart contracts ideal for implementing trustless systems where participants do not need to trust each other or a central authority.

The smart contract ensures transparency and accountability, while off-chain processing handles the computationally intensive tasks of the machine learning model. This hybrid approach balances security and efficiency, making the system practical for real-world use.

Ethereum was chosen for its mature ecosystem and developer support, though we also consider cost-effective alternatives like Binance Smart Chain and Polygon. These platforms offer lower transaction fees, making them suitable for high-frequency data submissions.

\subsection{Crowdsourcing Mechanism}

To encourage contributors to operate in good faith, we introduce a staking requirement whereby contributors must deposit a certain amount of currency before submitting data. The staking mechanism serves multiple purposes, including ensuring the smart contract balance and incentivizing contributors to provide high-quality submissions. The difference between the accuracy of the new and base models determines the weight assigned to each contributor. Based on this weight, the smart contract calculates and distributes rewards to the contributors. The contract retains the stake if no rewards are paid out.

\subsubsection{Staking}

Staking mechanisms, such as proof of stake, involve participants depositing a certain amount of cryptocurrency or tokens as collateral to ensure honest behavior~\cite{nguyen2019proof}. In our system, contributors must stake a certain amount of currency before submitting data. This stake serves as an incentive for contributors to provide high-quality submissions and a deterrent against malicious behavior. If the submitted data is verified to be accurate, contributors receive rewards. If not, they forfeit their stake. The stake requirement deters potential bad actors from exploiting the mechanism for undeserved rewards.

Furthermore, the smart contract is designed to operate on a blockchain network with low transaction fees, ensuring that users' stakes will always be greater than these fees. To further prevent any single user from monopolizing the incentives, we will limit the staking amount based on the total value of the contract. This means that no individual contributor can stake an excessively high amount and potentially withdraw all the funds from the contract. Additionally, the creator of the contract will stake an initial amount of funds to bootstrap the incentive pool, ensuring a steady flow of rewards in the early stages of the system. Over time, as contributors submit data that lowers the model's accuracy, whether they are malicious actors or well-intentioned users, they will lose their stakes. This loss of stakes will increase the funds within the smart contract, creating a self-sustaining source of incentives. This design ensures the system remains economically viable, even with frequent transactions.

\subsubsection{Incentives}

\begin{figure*}[t] \centering \includegraphics[width=\linewidth]{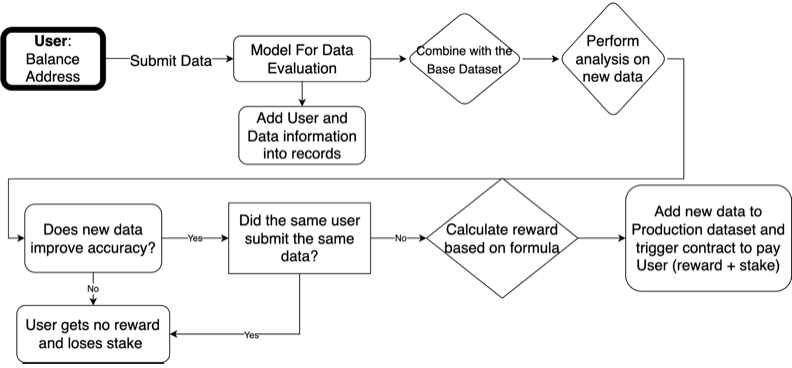} \caption{Incentive Mechanism Codified in the Smart Contract} \label{fig:incentive-mechanism} \end{figure*}

Our incentive mechanism, shown in \Cref{fig:incentive-mechanism}, uses smart contracts for transparency and security, evaluating contributor submissions against a base dataset to determine rewards based on model accuracy improvements. Contributors must submit their data and a \textbf{\textit{stake}} to the smart contract. The submitted data is then sent to the offline model trained on a base dataset. The objective is to evaluate whether the contributor's new data improves the model's accuracy on the base dataset. This evaluation is performed by calculating the accuracy of the current model on the base dataset and the accuracy of the new model trained on a combined dataset of existing data and new data. The \textit{weight} is then calculated as the difference between the accuracy of the models trained on the old and new dataset on the base dataset.

The \textbf{\textit{reward}} comprises the initial stake amount plus an additional reward proportional to the weight of the contributed data, incentivizing contributors to submit high-quality data that improve the model's performance. Contributors will not receive a reward for repeated data submissions.

This \textbf{\textit{reward}} is calculated as follows:

\begin{equation} reward = stake + (weight * stake) \end{equation}

Additionally, to discourage users from teaming up to scam the system, we included a mechanism to reduce rewards if the domain had already been seen multiple times in the system. For example, if the same domain were submitted three times, the weight of the reward would be adjusted by 1/3, i.e., \texttt{reward = stake + ((weight/3)*stake)}. These measures ensured that the incentive mechanism operated transparently and with integrity.

The smart contract handles all payouts, and users can access the improved model for a small fee for model inferencing purposes. The overall cost structure of the blockchain is designed to remain low, while malicious actors forfeiting their stakes will contribute to sustaining the system, ensuring that the benefits of improved spam detection outweigh the costs of maintaining the blockchain infrastructure.

\section{Evaluation}

We conducted a simulation with a popular web spam dataset to evaluate the end-to-end system performance,

\subsection{Dataset Description}

We used the phishing websites dataset\footnote{https://www.kaggle.com/datasets/aman9d/phishing-data}, which contains domain length, presence of dashes, and redirection patterns as features. 
We divided the dataset into three samples: base set, good actors' submissions, and bad actors' submissions.
The first sample, the base set, contained a reference of the entire dataset and performed at an accuracy of approximately 83\%. The second sample was for good actors, which had to perform at a higher accuracy than the base set, and the third sample was for bad actors, in which the labels were flipped to ensure poor-quality data submissions.

\subsection{Data Analysis}



\begin{figure}[t]
    \centering
    \begin{subfigure}{0.9\linewidth}
        \centering
        \includegraphics[width=\linewidth]{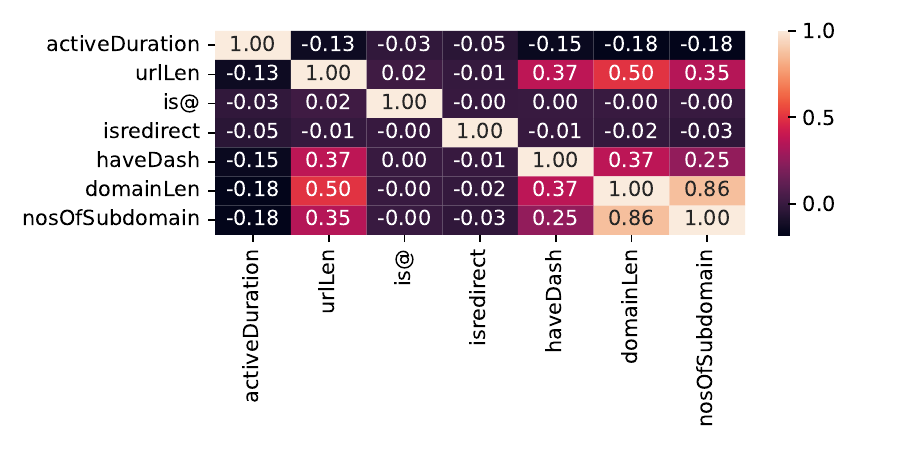}
        \caption{Feature Correlation}
        \label{fig:kaggle_feat_corr}
    \end{subfigure}

    \vspace{0.5cm}

    \begin{subfigure}{0.9\linewidth}
        \centering
        \includegraphics[width=\linewidth]{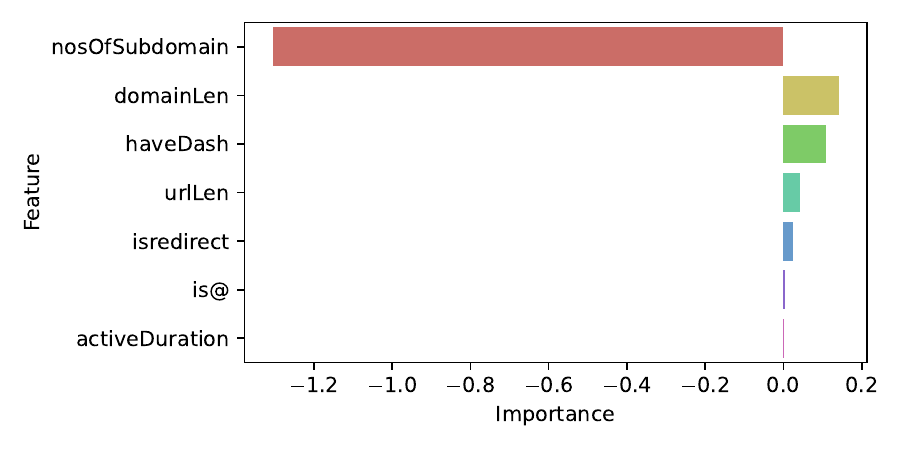}
        \caption{Feature Importance}
        \label{fig:kaggle_feat_imp}
    \end{subfigure}

    \caption{Feature Analysis}
    \label{fig:combined_features}
\end{figure}

The feature correlation heatmap in \Cref{fig:kaggle_feat_corr} highlights key relationships, such as the significant impact of URL patterns and domain length on spam detection, guiding feature selection for our model.
The feature importance scores in \Cref{fig:kaggle_feat_imp} show that features such as  \texttt{domainlength}, \texttt{hasdash} (in the domain), and \texttt{isredirect} (if it has a double dash, there is a chance it is a redirect) are some of the most important features to name a few. The \texttt{active duration} (obtained from the whois API) and \texttt{nosOfSubdomains} (in URL) have the most significant impact on predicting spam websites. Since most of these features (expect active duration) are based on URL patterns, we parsed new data inputs. 

\subsection{Simulation Setting}

\begin{figure*}[t]
    \centering
    \includegraphics[width=\textwidth]{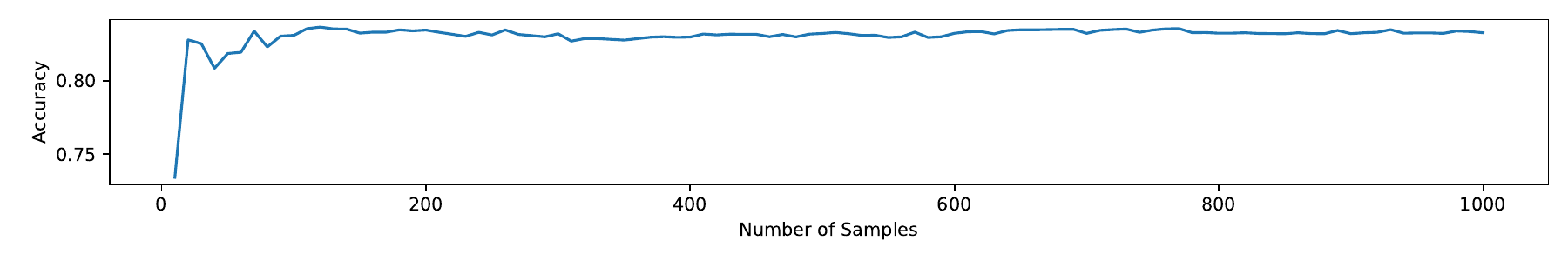}
    \caption{Model Evaluation With Increasing Dataset Sizes}
    \label{fig:kaggle_model}
\end{figure*}

To evaluate how the dataset will make predictions, we ran accuracy tests on increasing samples of 10 up to 1000 samples. From \Cref{fig:kaggle_model}, the model reached its maximum accuracy with tests of more than 200 samples. The results from this test provide evidence that these features identify spam vs. non-spam websites.


\begin{figure}[t]
    \centering
    \begin{subfigure}{0.45\linewidth}
        \centering
        \includegraphics[width=\linewidth]{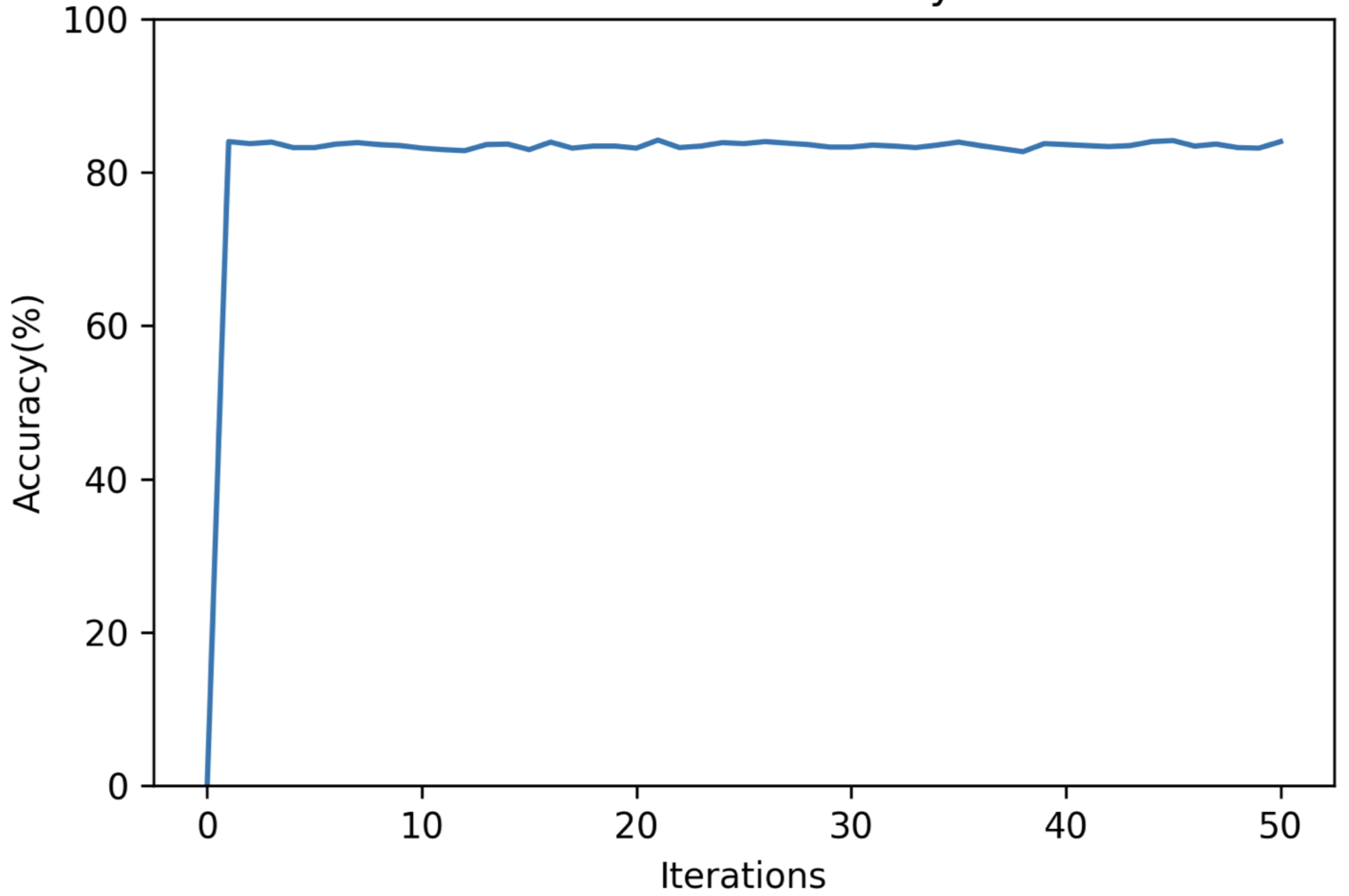}
        \caption{Simulated Balances}
        \label{fig:acc_overview}
    \end{subfigure}
    \hfill
    \begin{subfigure}{0.45\linewidth}
        \centering
        \includegraphics[width=\linewidth]{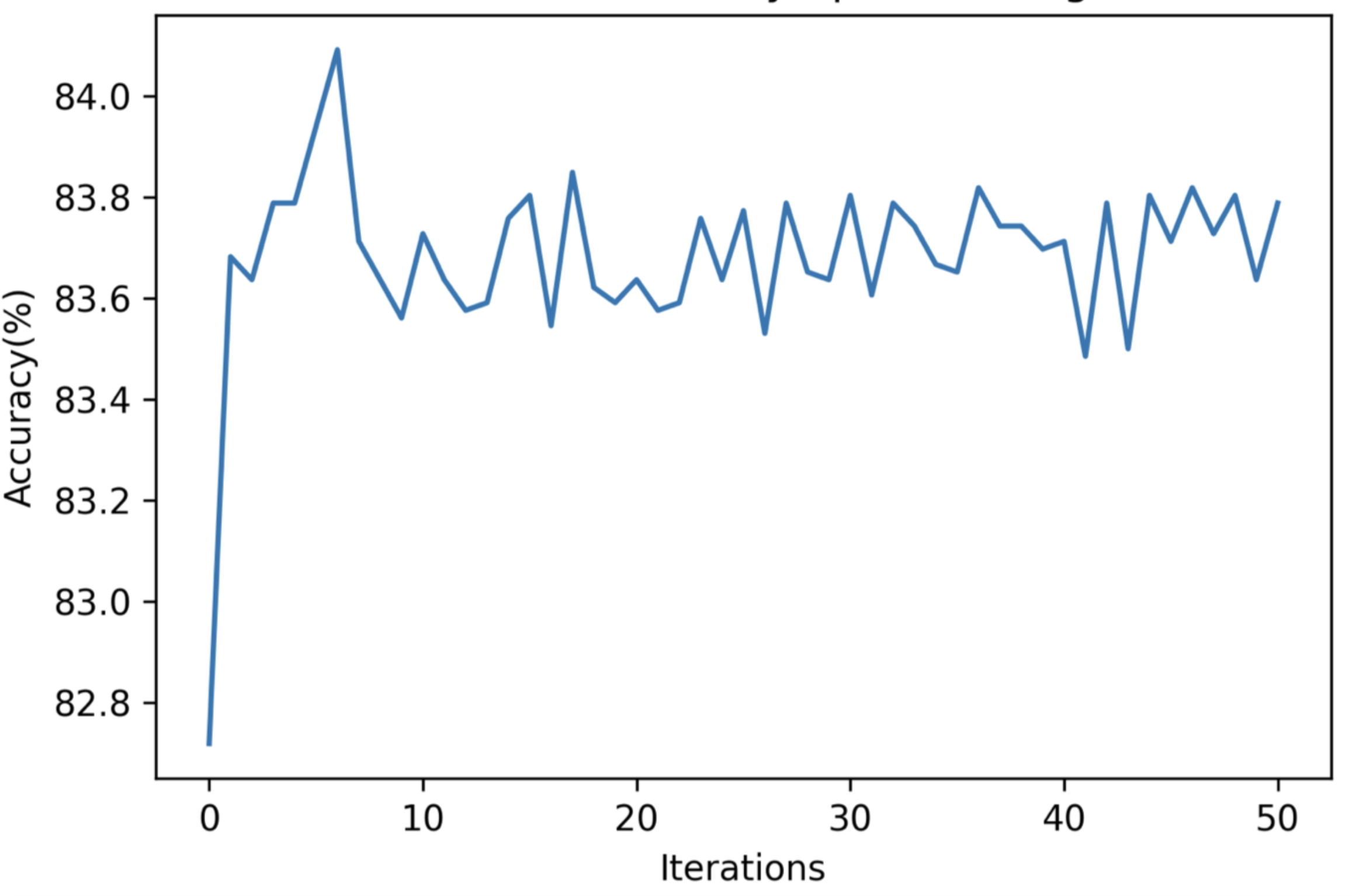}
        \caption{Accuracy}
        \label{fig:acc}
    \end{subfigure}

    \vspace{0.5cm}

    \begin{subfigure}{0.45\linewidth}
        \centering
        \includegraphics[width=\linewidth]{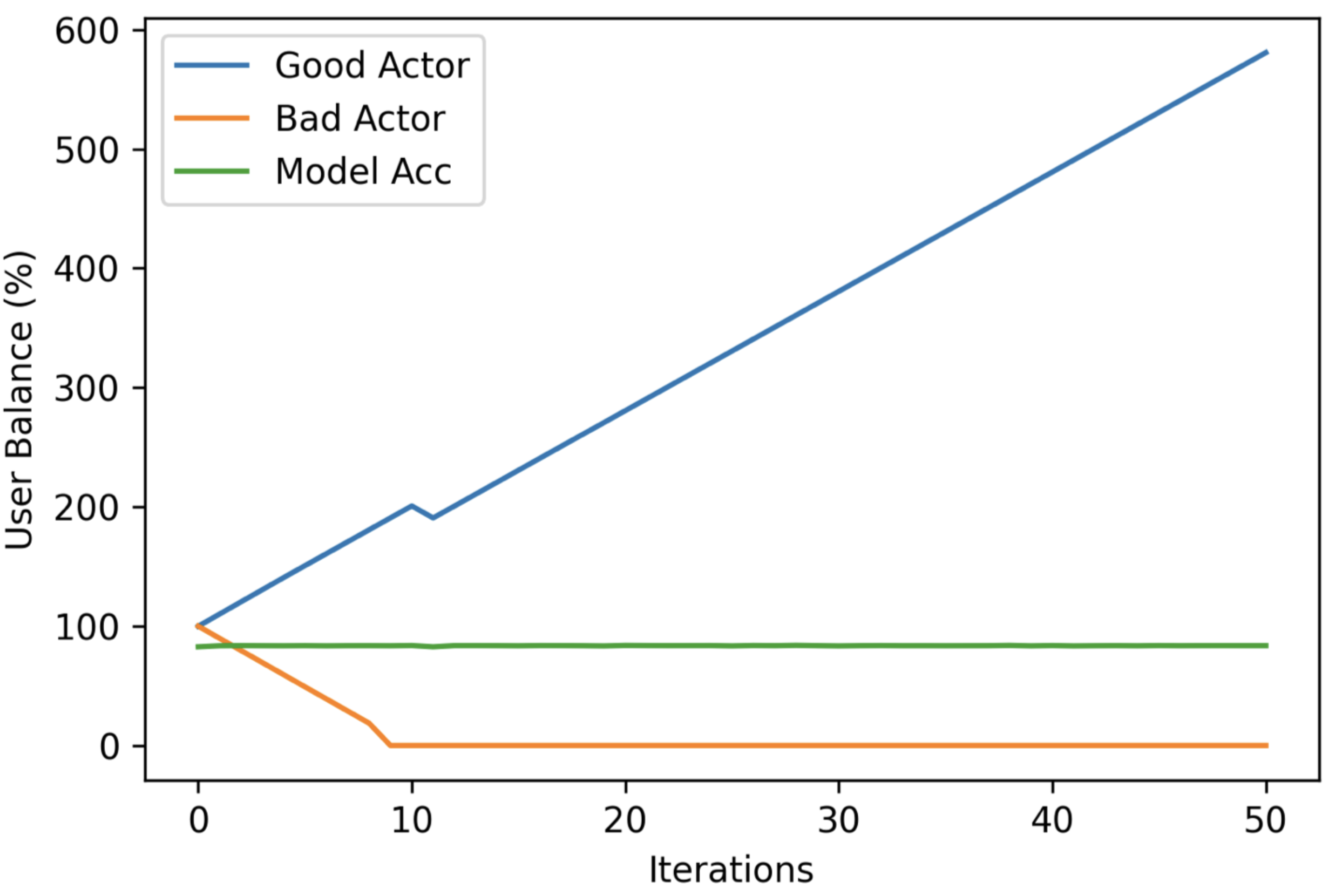}
        \caption{User Balance}
        \label{fig:sim}
    \end{subfigure}
    \hfill
    \begin{subfigure}{0.45\linewidth}
        \centering
        \includegraphics[width=\linewidth]{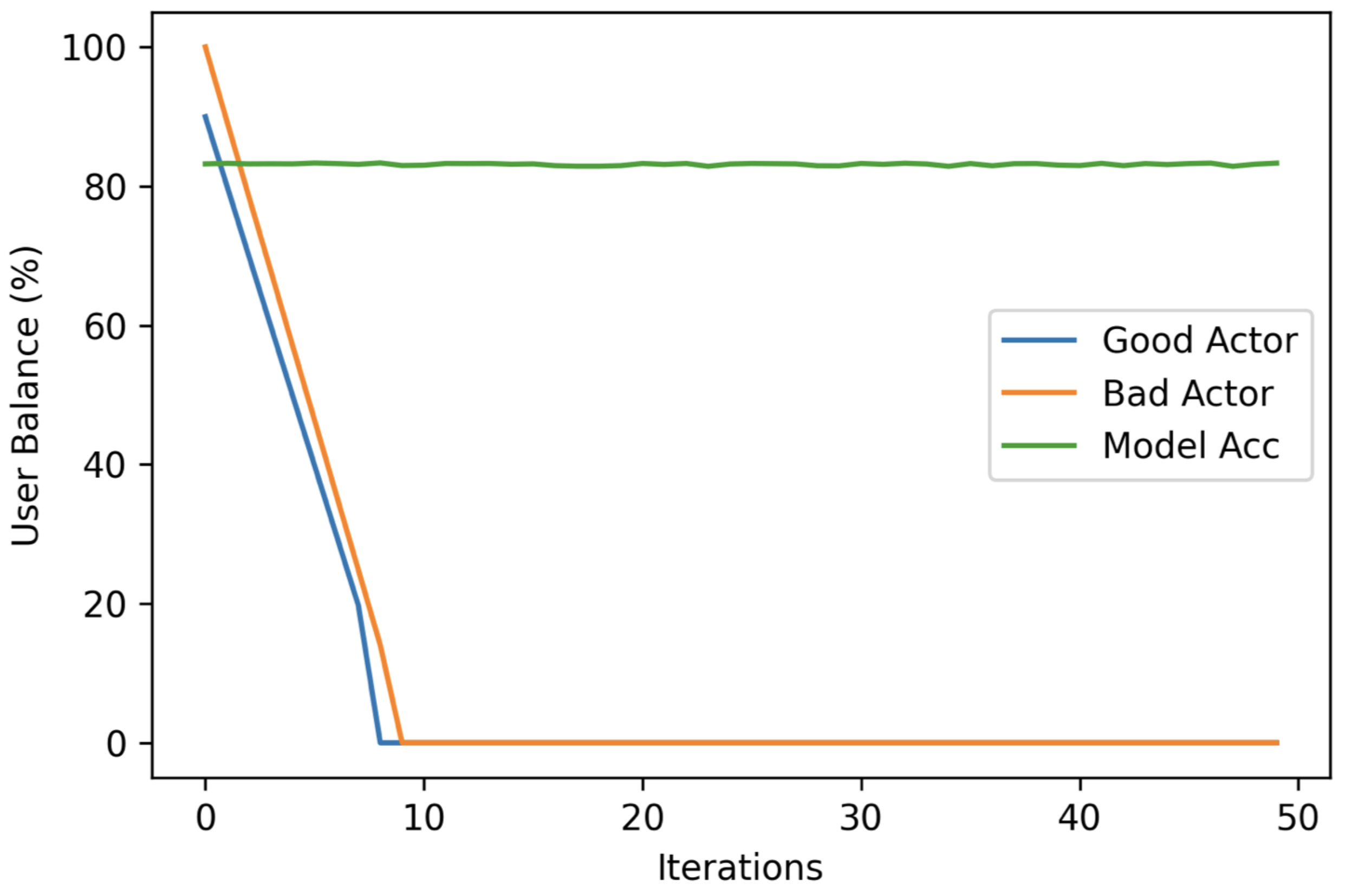}
        \caption{User Balance of Actor Submitting Same Data}
        \label{fig:bad}
    \end{subfigure}

    \caption{Incentive Mechanism Simulation Results}
    \label{fig:combined_figures}
\end{figure}

We then ran 50 iterations of users submitting data points to see how the balances and accuracy would perform. 
\Cref{fig:sim} illustrates the increasing balances of good actors, validating our incentive mechanism's effectiveness in promoting high-quality data submissions. 
Even if some of the submitted data were poor quality, good actors would still benefit from their contributions. As seen in \Cref{fig:acc} and \Cref{fig:acc_overview}, the accuracy of the production model will increase with the addition of the new "good" data.

\subsection{Deterring Bad Actors}
To deter bad actors from taking advantage of the system, we implemented measures to keep records of the data and which users submitted which data points. In \Cref{fig:bad}, we demonstrated that if a good actor attempted to repeatedly submit the same "good" data point to manipulate the system, they would lose their stake, which was done by maintaining a record of who submitted which data point in the form \texttt{[Contributor:[domain1,domain2,...]}, as conceptualized in \Cref{alg:weight}. Furthermore, we addressed the issue of users initially submitting high-quality data and then subtly introducing bad data by implementing a mechanism that does not accept data submissions if they lower the model's accuracy. Even if the actor has a high reputation, contributing data that degrades performance will lead to a loss of their stake and a decrease in their reputation. Additionally, we prevent collusion and redundant data flooding in two ways: First, if redundant data that has already been proven to improve accuracy is submitted, it does not change the user's weight, as the system has already seen and validated it. Second, while we acknowledge the potential for users to collude by upvoting each other's submissions, further testing will help identify patterns of collaboration. Nevertheless, submitting bad data—even with collaborators upvoting it—will not harm the model's accuracy, but we recognize the need to implement methods that prevent actors from financially benefiting from such behavior.

\section{Conclusion and Future Work}

Our blockchain-based incentivized crowdsourcing mechanism significantly enhances web spam detection accuracy by leveraging high-quality data contributions, setting a new standard for robust and reliable spam detection models to adapt to the ever-evolving landscape of spam websites and evasion techniques.
We implemented robust measures to prevent malicious behavior, such as staking requirements and reward adjustments for repeated data submissions, ensuring system integrity and fair participation.
The model's offline training and data evaluation reduce the risk of exposing sensitive information while providing a sustainable mechanism for leveraging blockchain technology's advantages.
Our mechanism paves the way for future advancements in web spam detection and other domains requiring high-quality crowdsourced data, promoting a safer and more trustworthy online environment.

Although our simulation results are promising, several limitations must be considered. Firstly, our dataset is limited to phishing websites, which may not represent all types of spam. Additionally, the simulation environment does not fully capture real-world dynamics, such as user interaction patterns and temporal changes in spam techniques. These limitations could affect the generalizability of our findings. Future studies should explore diverse datasets and real-world deployments to validate and extend our results.
Potential biases in our study include selection bias, as the dataset may not fully represent the variety of web spam encountered in different contexts. Additionally, the simulation environment may introduce biases not present in real-world applications. 
Several contextual factors may also influence the results. For instance, the evolving nature of spam tactics, regional differences in web usage, and varying levels of digital literacy could impact the effectiveness of our model. 
Therefore, to strengthen our findings further, future research endeavors should test the solution's robustness by incorporating data from multiple sources, exploring varying stake amounts, and conducting longitudinal evaluations to assess the mechanism's long-term effectiveness. 
Addressing privacy concerns and devising measures to prevent collusion among contributors are crucial aspects that warrant careful consideration to ensure the mechanism's fairness, reliability, and sustainability.

In conclusion, our mechanism stimulates collaboration, promotes continuous feature-set improvement, and encourages exploring various spam characteristics. By leveraging the collective intelligence and motivation of participants, our proposed approach holds the potential to overcome the challenges associated with limited resources and evolving spamming techniques, leading to more robust and effective spam detection models.


\bibliographystyle{splncs04}
\footnotesize{\bibliography{references}}
%
\end{document}